\documentclass[]{aastex631}
\providecommand{\tabularnewline}{\\}
\makeatother

\usepackage{babel}
\usepackage{graphicx}
\usepackage{appendix}
\usepackage{amsmath}
\usepackage{subfigure}

\shorttitle{Convolutional Neural Networks and Stokes Response Functions}
\shortauthors{Centeno et al.}

\graphicspath{{./}{figures/}}

\begin{document}

\title{Convolutional Neural Networks and Stokes Response Functions}

\author[0000-0002-1327-1278]{Rebecca Centeno}
\affiliation{High Altitude Observatory, National Center for Atmospheric Research, Boulder, CO}
\correspondingauthor{Rebecca Centeno}
\email{rce@ucar.edu}

\author[0000-0002-6544-5436]{Natasha Flyer}
\affiliation{Flyer Research LLC, Boulder,CO}

\author{Lipi Mukherjee}
\affiliation{High Altitude Observatory, National Center for Atmospheric Research, Boulder, CO}

\author[0000-0002-4996-0753]{Ricky Egeland}
\altaffiliation{now at NASA Johnson Space Center, Houston, TX.}
\affiliation{High Altitude Observatory, National Center for Atmospheric Research, Boulder, CO}

\author[0000-0001-6990-513X]{Roberto Casini}
\affiliation{High Altitude Observatory, National Center for Atmospheric Research, Boulder, CO}

\author{Tanaus\'u del Pino Alem\'an}
\affiliation{Instituto de Astrofísica de Canarias, E-38205, La Laguna, Tenerife, Spain}
\affiliation{Departamento de Astrofísica, Facultad de Física, Universidad de La Laguna, Tenerife, Spain}

\author{Matthias Rempel}
\affiliation{High Altitude Observatory, National Center for Atmospheric Research, Boulder, CO}

\begin{abstract}

In this work, we study the information content learned by a convolutional neural network (CNN) when trained to carry out the inverse mapping between a database of synthetic \ion{Ca}{2} intensity spectra and the vertical stratification of the temperature of the atmospheres used to generate such spectra. 
In particular, we evaluate the ability of the neural network to extract information about the sensitivity of the spectral line to temperature as a function of height.
By training the CNN on sufficiently narrow wavelength intervals across the \ion{Ca}{2} spectral profiles, we find that the error in the temperature prediction shows an inverse relationship to the response function of the spectral line to temperature, this is, different regions of the spectrum yield a better temperature prediction at their expected regions of formation. This work shows that the function that the CNN learns during the training process contains a physically-meaningful mapping between wavelength and atmospheric height.


\end{abstract}



\section{Introduction} \label{sec:intro}

In the rarefied environment of the solar chromosphere the radiation field is weakly coupled to the local thermodynamic conditions, and common approximations used to solve the polarized radiative transfer problem in the photosphere break down. The two main ones are 1) the assumption of local thermodynamical equilibrium (LTE), and 2) that the source of spectral line polarization is only due to the Zeeman effect. A more generalized approach needs to be adopted in the chromosphere, recognizing that the majority of spectral lines generated there form under non-LTE conditions, and in the presence of anisotropic excitation mechanisms that lead to the production of scattering polarization. As a result, the forward modeling of chromospheric spectral lines can be computationally very expensive, because of the need to solve the very complex problem of the statistical equilibrium of the atom under such conditions, and under the joint action of the Hanle and Zeeman effects.

The possibility to exploit spectropolarimetric observations of the solar atmosphere to infer its physical (thermodynamic and magnetic) state depends on our ability to understand and constrain a very complex and generally ill-posed inverse problem.
Spectral-line inversion codes for the photosphere have been developed and successfully employed for several decades thanks to the simplifications that are possible in the forward modeling of the formation of photospheric spectral lines. These codes carry out an ``inverse mapping'' from the observations to the physical state of the Sun’s atmosphere.
A traditional solution approach relies on non-linear least squares methods in order to minimize a merit function that quantifies the differences between the observed and modeled spectra. This entails solving the forward modeling of the spectral line many times until the inversion converges. In the case of chromospheric modeling, this approach adds yet another layer of iterations on top of an already computationally demanding problem. For this reason, spectral line inversion codes for chromospheric lines typically adopt different approximations to make the inversion problem tractable, by either relying on simplified atmospheres \citep{2008ApJ...683..542A} or assuming that the spectral line polarization is only due to the Zeeman effect \citep{2015A&A...577A...7S,2017SSRv..210..109D,2018A&A...617A..24M, 2019A&A...623A..74D}. Even after these simplifications, non-LTE inversion codes are still computationally demanding, have finicky initialization procedures, and often present unstable convergence behaviors, requiring a high level of experience to run them.

A different approach to spectral line inversions, which is based on machine learning (ML), is becoming very popular nowadays. While this approach is not completely new \citep[see e.g.][]{2001A&A...378..316C, 2005ApJ...621..545S}, it is gaining more traction in recent times because of the growing access to large computing power resources and to open-source ML software packages.  

By training a convolutional neural network (CNN) on the inversion results (and inversion fits) of synthetic spectra, \citet{2020A&A...644A.129M} were able to speed up the spectral line inversion of these data by 5 orders of magnitude, compared to the performance of a traditional Levenberg-Marquardt-based inversion scheme.
Following a slightly different philosophy, \citet{2019A&A...626A.102A} trained two different CNN architectures on synthetic spectropolarimetric data from radiative magneto-hydrodynamical MURaM simulations. Instead of doing this on a pixel-by-pixel basis, these authors fed the CNNs 2-dimensional patches of the simulated Sun's surface, taking advantage of the horizontal continuity of the data. The CNNs so constructed are able to retrieve the thermodynamical and magnetic properties of the photosphere as a function of geometric height, as well as to deconvolve the data from the instrument's point spread function. This technique was successfully applied to Hinode/SP observations and the authors predict that it could invert all the data from the entire mission in just a matter of minutes.
A recent work by \citet{2021arXiv210812421H} shows that ML techniques can also be used to cross-calibrate spectral line inversion results from instruments that observe the Sun in fundamentally different ways (in their case, Hinode/SP and SDO/HMI). 
\citet{2019ApJ...875L..18S} apply deep learning techniques to the inversion of chromospheric intensity spectra from the IRIS mission, resulting in a decrease of 5 to 6 orders of magnitude in the computing time for the retrieval of the thermodynamical properties of the chromosphere. While the accuracy of the results is often reported to be good in a qualitative manner, \cite{2021arXiv210309651G} claim that these kinds of approaches can be used to find suitable guess atmospheres to feed to traditional non-linear least squares inversion codes, resulting in a significant reduction in the number of inversion cycles, and cutting the computing time by a factor of 2 to 4 .

An obvious limitation of the NN approach to spectropolarimetric inversions is that the atmospheric and magnetic field models that can be inferred through data inversion are restricted to those contained in the physical parameter space adopted for the NN training (i.e., NNs cannot be used to extrapolate solutions outside of the training set). Contrary to how traditional inversion codes work, the NN inversion does not call an underlying radiative transfer model. Given enough data, however, it is able to reconstruct a function that carries out an approximate mapping between the Stokes profiles at the atmospheric properties that lead to them. This begs the question of whether or not this mapping function is able to glean physically-meaningful information from the databases on which it has been trained.

In this work, we conduct a detailed study of the information content that a CNN is able to infer when trained to carry out the inverse mapping between a database of synthetic intensity spectra and the atmospheric temperature profiles used to generate them. In particular,
we analyze what the CNN is able to learn about the sensitivity to temperature of the two lines of \ion{Ca}{2} at 3934 \AA\ (K-line) and 8542 \AA\ (IR2), as a function of wavelength and atmospheric height. This sensitivity is derived from the errors in the temperature prediction when the CNN is trained to carry out the inverse mapping on narrow spectral windows within each of the spectral lines.
In Section \ref{sec:spectra}, we describe the numerical setups used to create the databases of atmospheric models and \ion{Ca}{2} spectra. Section \ref{sec:CNN_arch} presents the CNN architectures used for this work, and describes the methodology that we follow to quantify the information content learned by the CNN. A comparison of these results with the response functions of \ion{Ca}{2} K and IR2 to temperature is presented in Section \ref{sec:rf}, followed by our concluding remarks and an outlook for future developments.

\section{Creation of databases of atmospheric models and spectra}\label{sec:spectra}

In order to compile a sufficiently large database of 1-dimensional (1D) model atmospheres and spectra on which to train the CNN, we used a snapshot of a 3D radiative magneto-hydrodynamic simulation from the MURaM code. In particular, we chose a Quiet Sun (QS) simulation  \citep{2014ApJ...789..132R}, with a horizontal spatial domain of 24.5x24.5 Mm${^2}$ and a vertical extent of 8.2 Mm ($\sim$2\,Mm of which are above the solar surface). The spatial grid is sampled with a 16 km spacing in all directions. The simulation was started from thermally relaxed, non-magnetic convection; successively, a constant, vertical, seed magnetic field of $10^{-3}$\,G was added, unevenly distributed over the horizontal plane. Once the simulation reached the saturation regime the average magnetic field strength was 60 G, while the average net vertical flux was negligible. 

While this particular MURaM simulation does not produce realistic physical conditions in the solar chromosphere, the main goal of this work is to understand the information content learned by the CNN from its training dataset. The methodology presented here can later be utilized on increasingly realistic numerical simulations, with the ultimate objective of developing a functional ML spectral line inversion code that can be applied to solar observations. 

The database of 1D model atmospheres was compiled by extracting individual columns from the 3D MURaM cube. In order avoid redundancy and correlation in our atmospheric database, we selected columns on a sparse horizontal grid (keeping one in every 8, in each spatial direction).

In order to generate Stokes spectra from the different realizations of the MURaM atmospheres we use the Hanle-RT code \cite[][]{2016ApJ...830L..24D}. Hanle-RT is a polarized radiative transfer code for 1-D (plane-parallel) atmospheres. It accepts any multi-term atom (LS-coupling scheme) without hyperfine structure,\footnote{Multi-level atoms with hyperfine structure can also be modeled with this code.} and in the presence of magnetic fields of arbitrary strength. The code takes into account the effects of non-isotropic radiative excitation of the atomic system, which are responsible for the manifestation of scattering polarization. Therefore, the code can model seamlessly all possible regimes of magnetic-induced effects on the polarization of spectral lines: the Hanle, Zeeman, and (incomplete) Paschen-Back effects, as well as the polarization effects associated with level crossing and anti-crossing interference. The code fully implements the effects of inelastic, elastic, and depolarizing collisions, and also includes a comprehensive description of partially coherent scattering (partial redistribution, or PRD) from polarized atoms \citep[limited to 3-term atoms of the $\Lambda$-type; see][]{2016ApJ...824..135C}, which is important for modeling deep chromospheric lines.

The atomic model of \ion{Ca}{2} used to solve the statistical equilibrium (SE) includes the five bound levels of the lowest three terms of the ion, as well as the \ion{Ca}{3} ground state. Such a model describes five bound-bound radiative transitions, namely, the \ion{Ca}{2} H \& K lines in the near UV and the \ion{Ca}{2} IR triplet. The IR triplet is computed assuming complete frequency redistribution (CRD), whilst the H \& K lines are treated in PRD. Cross-distribution (Raman scattering) is allowed between these sets of transitions. For computational expediency, the system is treated as a 5-level system, rather than the actual 3-term atom. Thus, the effects of $J$-level quantum interference (incomplete Paschen-Back effect, and level crossing and anti-crossing) are not taken into account in this test model, but they would have a negligible impact on the Stokes I profile anyway.

The MURaM simulation provides the thermodynamical and magnetic quantities as a function of geometric height needed to compute the Stokes spectra: temperature, gas pressure, line-of-sight (LOS) velocity, and magnetic field vector. Often times, microturbulent velocity is added to the models for broadening purposes, in order to increase the similarity between synthetic spectra and observations \citep{2012A&A...543A..34D}. A direct comparison with observations, however, is out of the scope of this work, so we do not see the need to artificially increase the complexity of the model for this sake. 

\begin{figure}[ht!]
\centering
\includegraphics[width=0.6\textwidth]{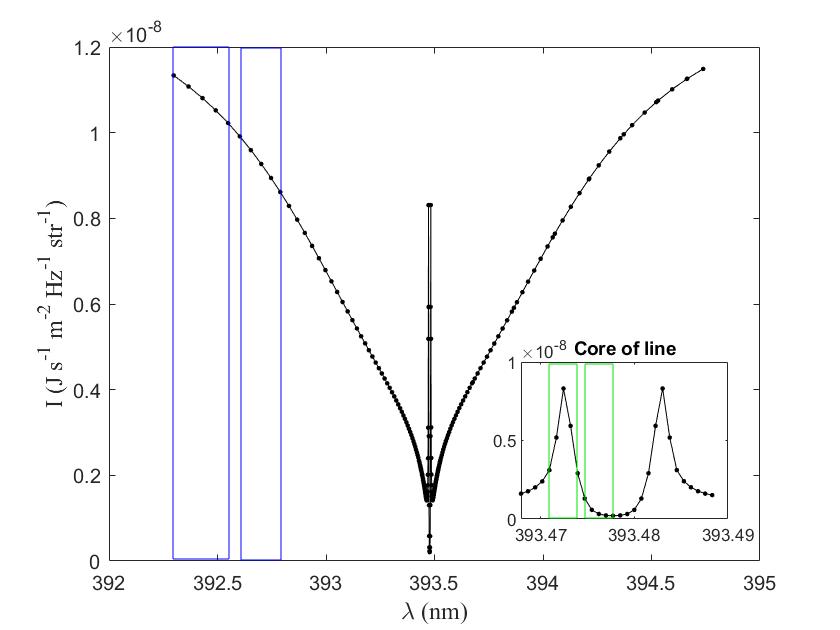}
\caption{An example of a \ion{Ca}{2} K Stokes I profile as a function of wavelength. The black dots represent the wavelength sampling across the line, which is non-equidistant. The figure inset zooms into the core of the line. The vertical blue and green boxes showcase a sliding wavelength window, covering five wavelength sampling points, used to analyze the sensitivity of the spectral line to temperature as a function of height (described in Section \ref{sec:CNN_arch}) .\label{fig:caii}}
\end{figure}

Before feeding these quantities to Hanle-RT, we set the LOS velocity to 0 km/s in the interest of preserving the (non-linear) mapping between absolute wavelength and atmospheric height. If velocities are present, this mapping is subject to further smearing due to the Doppler shifting of the spectral lines\footnote{The reader interested in the accuracy of the temperature prediction using CNNs when velocities are included in the spectral line synthesis, is referred to section \ref{appendix:velocities} of the Appendix.}. The 1D thermodynamical profiles are then smoothed in the vertical direction by applying a Savitzky-Golay filter \citep[][]{savgolay} with a 1st-order polynomial and a window size of 17 grid points ($\equiv 256$~km). Then, they are re-sampled onto a 32 km grid, and the portion of the model that lies below -192 km is discarded (a portion of atmosphere below 0 km is retained \footnote{height $= 0$ km is defined as the point where the mean continuum optical depth at 500 nm in the MURaM cube is unity ($\tau_{500} = 1$).} in order to maintain its small contribution to the formation of the wings of the spectral lines). 
This vertical smoothing was carried out in order to reduce the LOS gradients and the inherent fluctuations in the model atmosphere, while retaining the main trends and the variability across the different realizations in the atmospheric database. This process yields 36,828 unique 1D model atmospheres, that span from $-$192\,km to 1888\,km in height, sampled in steps of 32 km. These model atmospheres are given to Hanle-RT as an input. Hanle-RT then calculates the Stokes profiles for the 5 \ion{Ca}{2} lines contained in the model atom, assuming an observing geometry of $\mu=1.0$ (where $\mu$ is the cosine of the heliocentric angle). Figure \ref{fig:caii} shows an example of a \ion{Ca}{2} K intensity profile as a function of wavelength. The wavelength sampling is non-equidistant, being much denser in the core of the line than in the wings (the former is enlarged in the inset of the figure for better viewing). This is necessary in order to ensure numerical stability while solving the radiative transfer because of the steeper gradients in the core of the line compared to the wings. The blue and green boxes delimit narrow (5 grid-point) wavelength windows, the purpose of which will be discussed in the next section.

\begin{figure}[ht!]
\centering
\includegraphics[width=0.6\textwidth]{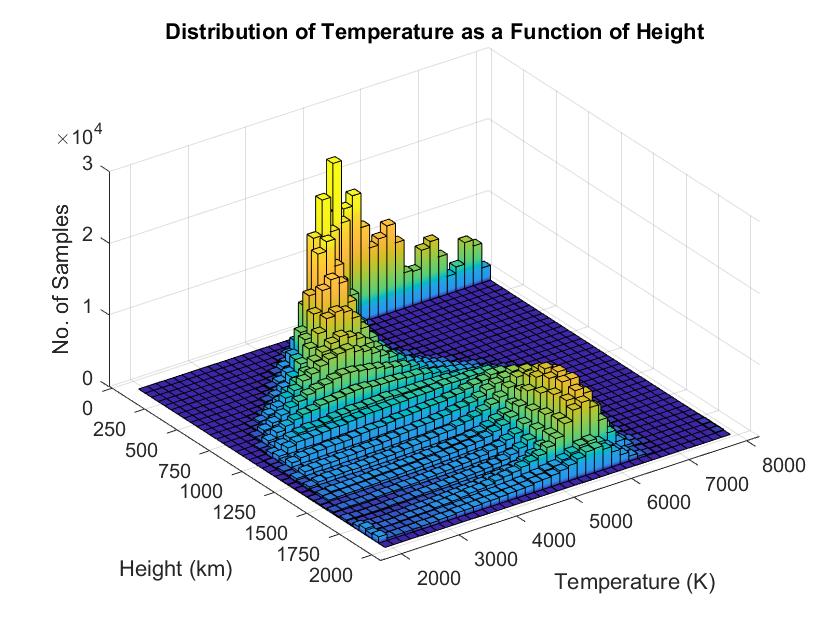}
\caption{3D histogram of the distribution of temperature as a function of height for the 36,828 samples in the database.\label{fig:3DhistT}}
\end{figure}

In this way, we create two databases, one of model atmospheres and one of Stokes spectra, with a known one-to-one correspondence between each individual 1D physical state and each Stokes vector. We can then use these databases to train a neural network to learn the mapping between the synthetic ``observations'' (spectra) and the physical conditions in the Sun’s atmosphere. 
In what follows, we will map the Stokes I profiles of the \ion{Ca}{2} K and IR2 lines (separately) to the atmospheric temperature as a function of height using the regression CNN outlined in the next section.
  The distribution of temperature as function of height for the 36,828 samples in our atmospheric database is shown in Figure \ref{fig:3DhistT}. In the photoshpere ($\sim$0-500\,km), the temperature samples are concentrated in a narrow range of values (4000\,K to 6500\,K), while as we approach the chromosphere, the distribution spreads, showing a much larger range of temperatures (typically between 2000\,K and 7000\,K). As we will see later, this will affect the accuracy of the CNN temperature retrieval, with the lowest errors found below 500\,km.


\section{CNN architecture}\label{sec:CNN_arch}
The regression CNN developed for this study is programmed in Keras \citep{chollet2015keras} (see Appendix for the code).  As shown in Table \ref{tbl:CNN}, it is a 9-layer CNN with 3 convolutional layers, each followed by an Exponential Linear Unit (ELU) 
non-linear activation function with its hyperparameter set to 1.
The choice of activation function relies on the faster learning speeds and higher accuracy that the ELU confers to deep NNs \citep[][]{clevert2016fast}. A mean absolute percentage error ($100/N\sum_{i=1}^N |true_i - predicted_i|/|true_i|$) loss function is used to calculate the model error from the 60 unit output layer (corresponding to the temperature sampled at 60 geometric heights) during the optimization process. The optimization algorithm used is Adam \citep{kingma2017adam}, a stochastic first-order gradient method with the learning rate set to 0.01. The batch size is 50, with $67\%$ of the data used for training, $3\%$ for validation, and $30\%$ for testing. The network's batch size was kept optimally small to achieve higher accuracy without capturing noise \citep[as illustrated in][]{keskar2017largebatch}. The CNN was trained for 150 epochs. A min-max scaling was used on the input data (Stokes I) to normalize it between 0 and 1 \citep{minmax}.

\begin{table}
\centering
\begin{tabular}{|c|c|c|c|} 
\hline
Layer Type & Description & 
$\begin{array}{c}
\textrm{Activations/}\\
\textrm{Output Shape}
\end{array}$ & 
$\begin{array}{c}
\textrm{Learnable}\\
\textrm{Parameters}
\end{array}$\tabularnewline
\hline 
\hline 
image\_input & 253\texttimes 1 image (Whole Spectral Line) &  & ---\tabularnewline
\hline 
 \multicolumn{4}{|c|}{} \tabularnewline
\hline 
Convolution  & 32 3\texttimes 1 convolutions, stride {[}1 1{]} & 253\texttimes 1\texttimes 32 & $\begin{array}{c}
\textrm{Weights}:(3\times1)\times32\\
\textrm{Bias}:32\times1
\end{array}$\tabularnewline
\hline 
ELu & ELU nonlinear activation function & --- & ---\tabularnewline
\hline 
 \multicolumn{4}{|c|}{} \tabularnewline
\hline 
Convolution & 64 3\texttimes 1 convolutions, stride {[}1 1{]} & 253\texttimes 1\texttimes 64 & $\begin{array}{c}
\textrm{Weights}:(3\times32)\times64\\
\textrm{Bias}:64\times1
\end{array}$\tabularnewline
\hline 
ELu & ELU nonlinear activation function & --- & ---\tabularnewline
\hline 
 \multicolumn{4}{|c|}{} \tabularnewline
\hline 
Convolution & 64 3\texttimes 1 convolutions, stride {[}3 1{]} & 85\texttimes 1\texttimes 64 & $\begin{array}{c}
\textrm{Weights}:(3\times64)\times64\\
\textrm{Bias}:64\times1
\end{array}$\tabularnewline
\hline 
ELu & ELU nonlinear activation function & --- & ---\tabularnewline
\hline 
 \multicolumn{4}{|c|}{} \tabularnewline
\hline 
Dense & $\begin{array}{c}
60\textrm{ unit fully connected output layer}\\
\textrm{(Temperature output at 60 heights)}
\end{array}$  & 60\texttimes 1 & $\begin{array}{c}
\textrm{Weights}:(85\times64)\times60\\
\textrm{Bias}:60\times1
\end{array}$\tabularnewline
\hline 
Loss\_MAPE & Mean absolute percentage error  & --- & ---\tabularnewline
\hline
\end{tabular}
\caption{\label{tbl:CNN}CNN architecture for training the \ion{Ca}{2} K spectral line (253 values of the intensity) to map the temperature at 60 geometric heights.}
\end{table}

Of special note to this CNN architecture is that no pooling layers are used to reduce the spatial size of the data. Such layers usually occur in CNN architectures to downsample the feature maps that are inputted into the next convolutional layer and thus hopefully retain the most important features in a map while decreasing the computation in the network. In general, this is usually warranted when attempting to do classification on high-dimensional images (e.g. reducing an 128 by 128 image to a vector of 10 for classification through a softmax function). However, in the presented problem we are doing 1D convolutions on a vector of length 253 (wavelength samples) with an output of length 60 (geometric heights), in an attempt to map the spectral wavelength domain to the temperature-height domain. Instead of pooling layers, we used a {\em striding} approach to perform downsampling in the last convolution layer. By increasing the stride length of the convolutional filter we can still reduce the spatial size of the output (layer output = (input - filter size)/stride + 1) while the convolutional filter weights benefit from {\em ``learning''} on a larger input feature map than would have occurred if a pooling layer had been used instead.

\begin{figure}[t!]
\centering
\includegraphics[width=0.45\textwidth]{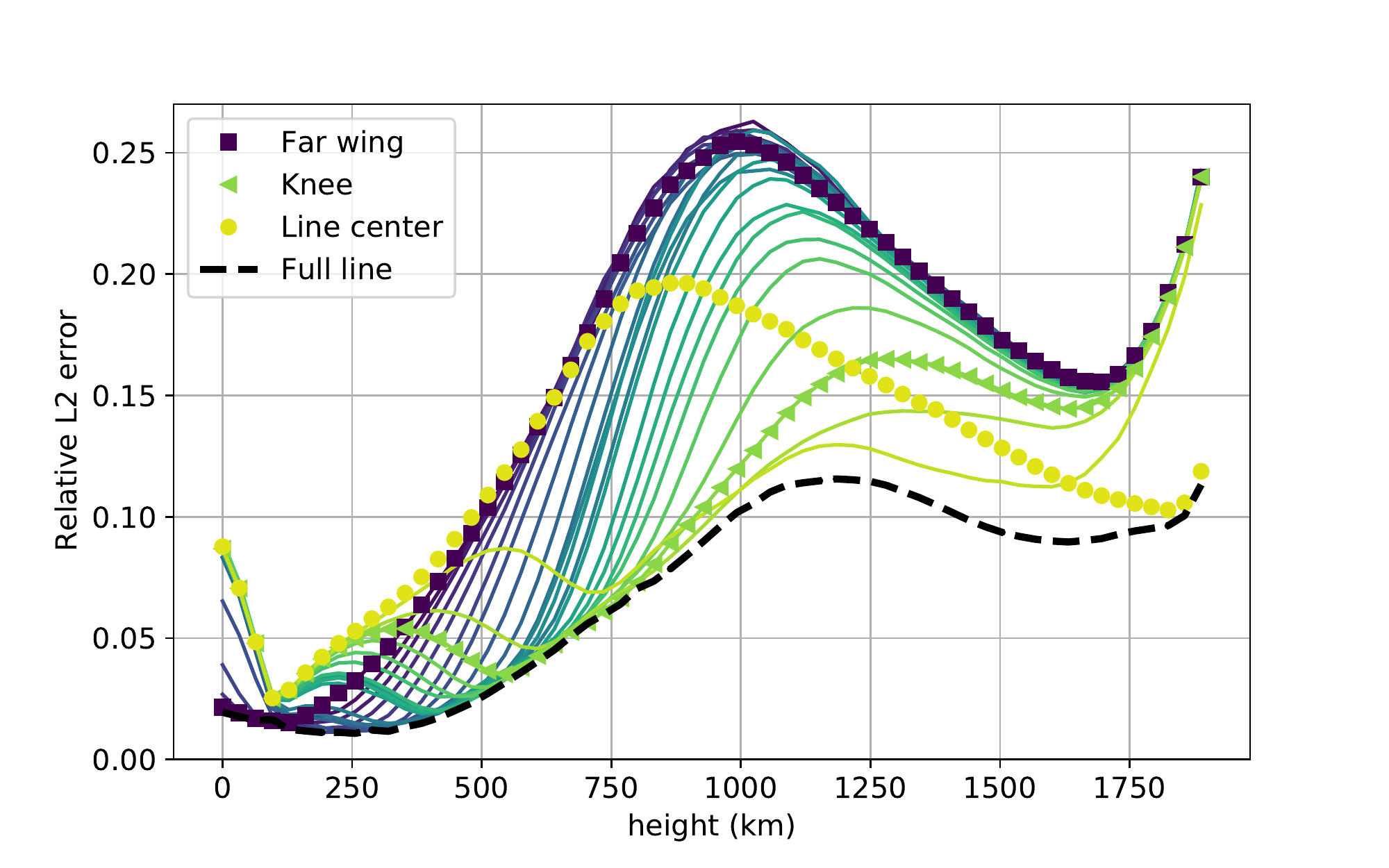}
\includegraphics[width=0.45\textwidth]{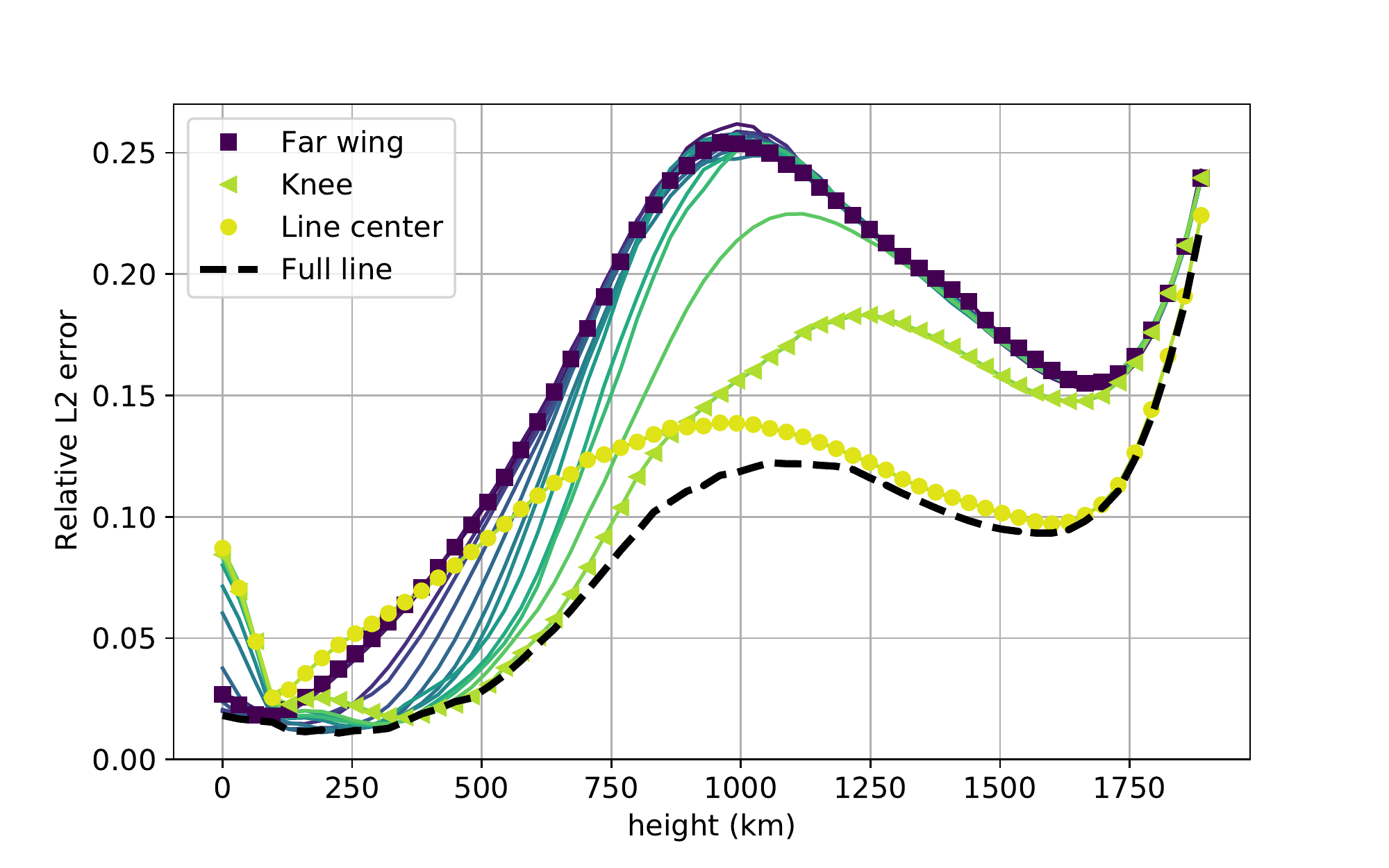}
\caption{CNN-derived sensitivity of the spectral line. The results for \ion{Ca}{2} K are shown on the left and those for \ion{Ca}{2} 8542 \AA\ are shown on the right. The colored lines show the L2-error in the temperature prediction when the CNN is trained on short spectral windows along the spectral line. Dark blue/purple corresponds to the case when the window is in the far wing of the line and lightest shade of green corresponds to its placement in the core. The black dashed line corresponds to the error in the temperature prediction when training the CNN on the whole spectral line at once. 
\label{fig:cnnrf}}
\end{figure}

This CNN architecture is trained on the \ion{Ca}{2} K and IR2 intensity spectra separately, to carry out the inverse mapping from the spectral profiles to the atmospheric temperature as a function of height. The black dashed lines in Fig. \ref{fig:cnnrf} show the CNN prediction uncertainty, as a function of height, for the the CNN trained on \ion{Ca}{2} K (left) and IR2 (right). The errors are given in terms of the L2-norm:

\begin{equation}
\text{L2}  = \sqrt{\frac{\sum_{i=1}^N (t_i - p_i)^2}{\sum_{i=1}^N t_i^2}} 
\label{L2}
\end{equation}

\noindent where $t$ and $p$ represent the true and predicted temperatures, respectively, and the summation is carried out over all the samples in the database. The L2-norm is very similar to the root mean square error, but the normalization is done with respect to the true vector as opposed to the number of samples, $N$. The advantage is that it gives a relative error that is dimensionless. 

\begin{figure}[ht!]
\centering
\includegraphics[width=0.9
\textwidth]{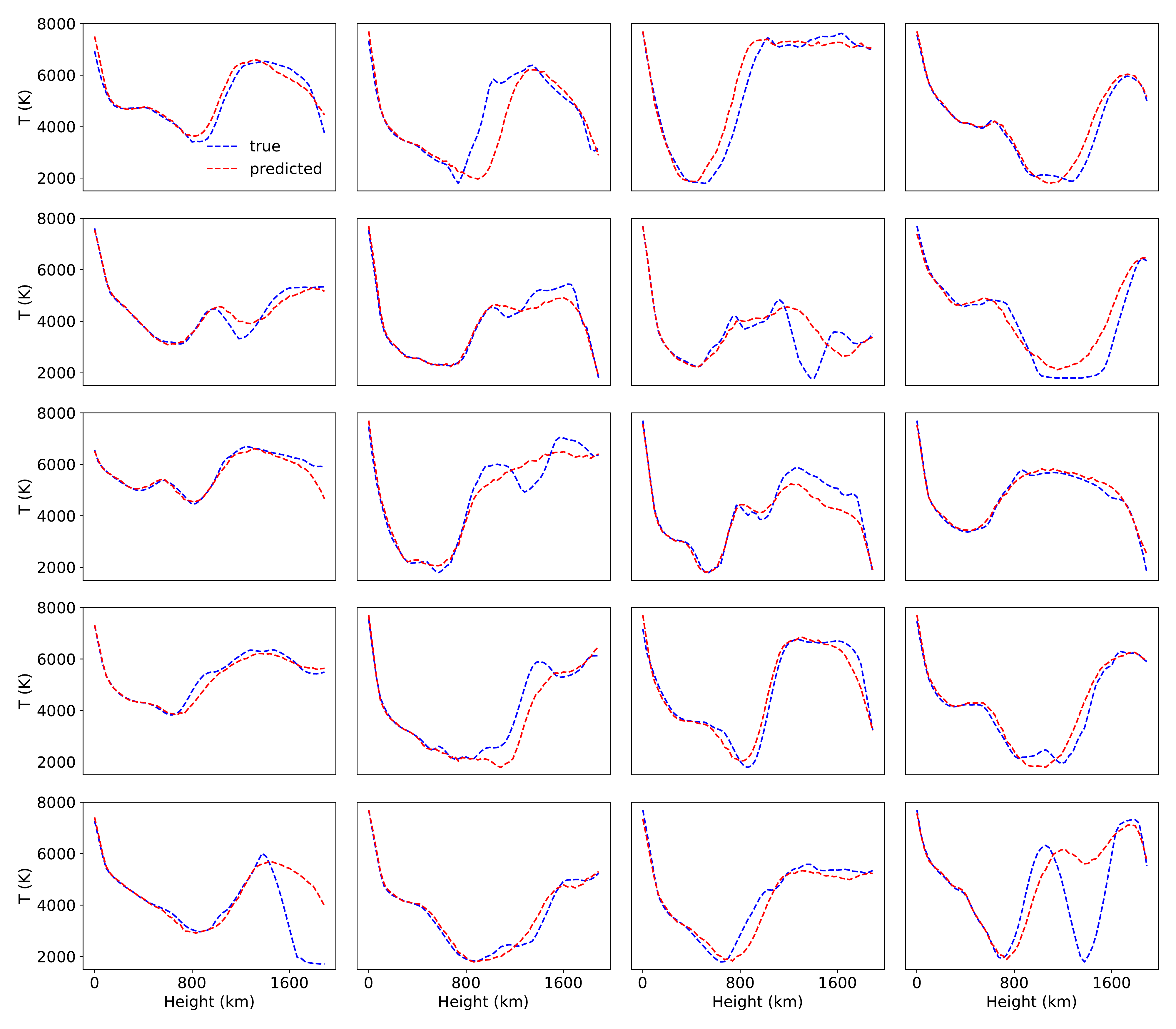}
\caption{True (blue) vs. predicted (red) temperature as a function of height for a selection of 25 random samples from the testing set. \label{fig:RandomT}}
\end{figure}

The results for the \ion{Ca}{2} K and IR2 lines are strikingly similar. As hinted earlier, the error in the temperature prediction is smaller in the lower parts of the atmosphere (for both spectral lines) than it is at higher layers. Interestingly, the error in the temperature prediction at the very top of the atmosphere (above $\sim 1700$ km) is much larger when the CNN is trained on IR2 than when it is trained on the K-line. This is likely due to the fact that the core of the latter forms higher in the atmosphere \citep[see, for instance][]{Leenaarts2013}, and therefore retains sensitivity for mapping the temperature at these heights. An analysis of (and comparison with) the Stokes response function to temperature for both of these lines is presented in section \ref{sec:rf}. Figure \ref{fig:RandomT} shows examples of the true (blue) vs. predicted (red) temperature stratification for a random selection of samples in the testing set. While the CNN is not able to capture the small-scale variations in the temperature, it predicts its overall stratification with strikingly good accuracy.

\subsection{CNN setup for mapping spectral line sensitivity to temperature at different heights}

Given enough data, the CNN-based inversion is able to reconstruct a function that carries out an approximate mapping between the spectral line intensity and the temperature stratification. Throughout the remainder of this manuscript we will try to quantify how much information this CNN inversion method is able to glean about the sensitivity of different parts of the spectral line to different heights in the atmosphere, and whether or not this bears any relation to the temperature response function of the spectral line. In other words, we are interested in finding out if there is physically-meaningful information in the mapping function between the wavelength and the height domains that the CNN learned during its training.

To analyze how different wavelengths of the \ion{Ca}{2} K and IR2 spectral lines are sensitive to the temperature at different atmospheric heights, we train a two-layer CNN (given in Table \ref{tbl:CNN_1layer}) on a short wavelength window composed of five consecutive wavelength samples in the wing of Stokes I (such a wavelength window is represented by the left-most blue box in Fig. \ref{fig:caii}). The window, always of five wavelength samples, is then slid across the spectral line with a stride of 5 wavelength positions and the CNN is trained again on the new wavelength range (indicated by the second blue box in Fig. \ref{fig:caii}). This process is repeated all the way to the central wavelength of the spectral line (right-most green box in the inset of Fig. \ref{fig:caii}), and the  L2 errors for the temperature prediction as a function of height are computed for each position of the short spectral window. Only half of the spectral line is considered because of the (near perfect) symmetry properties of Stokes I in the absence of velocity gradients. Notice that the window width (in wavelength units) narrows as it moves from the wing to the core of the line due to the non-equidistant wavelength sampling. We decided to work with this uneven sampling because the information density in the core of the line is much larger than in the wings.  

\begin{table}
\centering
\begin{tabular}{|c|c|c|c|} 
\hline
Layer Type & Description & 
$\begin{array}{c}
\textrm{Activations/}\\
\textrm{Output Shape}
\end{array}$ & 
$\begin{array}{c}
\textrm{Learnable}\\
\textrm{Parameters}
\end{array}$\tabularnewline
\hline 
\hline 
image\_input & 5\texttimes 1 image (Window of Spectral Line) &  & ---\tabularnewline
\hline 
 \multicolumn{4}{|c|}{} \tabularnewline
\hline 
Convolution  & 32 3\texttimes 1 convolutions, stride {[}1 1{]} & 5\texttimes 1\texttimes 32 & $\begin{array}{c}
\textrm{Weights}:(3\times1)\times32\\
\textrm{Bias}:32\times1
\end{array}$\tabularnewline
\hline
ELu & ELU nonlinear activation function & --- & ---\tabularnewline
\hline 
Convolution  & 64 3\texttimes 1 convolutions, stride {[}1 1{]} & 5\texttimes 1\texttimes 64 & $\begin{array}{c}
\textrm{Weights}:(3\times1)\times64\\
\textrm{Bias}:64\times1
\end{array}$\tabularnewline
\hline
ELu & ELU nonlinear activation function & --- & ---\tabularnewline
\hline
\multicolumn{4}{|c|}{} \tabularnewline
\hline 
Dense & $\begin{array}{c}
60\textrm{ unit fully connected output layer}\\
\textrm{(Temperature output at 60 heights)}
\end{array}$  & 60\texttimes 1 & $\begin{array}{c}
\textrm{Weights}:(5\times32)\times60\\
\textrm{Bias}:60\times1
\end{array}$\tabularnewline
\hline 
Loss\_MAPE & Mean absolute percentage error  & --- & ---\tabularnewline
\hline
\end{tabular}
\caption{\label{tbl:CNN_1layer} Two-layer CNN for training on a short wavelength window within the spectral line to carry out the mapping between the intensity and the temperature as a function of height.}
\end{table}


 The colored lines in Fig. \ref{fig:cnnrf} show the L2 prediction errors derived from training the two-layer CNN on these short spectral windows, for \ion{Ca}{2} K (left) and for IR2 (right).
The hue of the lines in the plots reflects the positioning of the 5-sample wavelength windows along the spectral line, with dark blue corresponding to the far wing and light green to the line core (the results for specific wavelength windows are highlighted by different symbols in the plots).

Examine the results for the K-line (left). At the lowest heights ($<160$ km), the best prediction is achieved by the darkest blue line (also represented with squares), which corresponds to the errors of the CNN trained in the far wing of the spectral line (more than 1.2 nm away from the line center, $\lambda_0 $). 
As the spectral window is slid inwards (and the color of the lines in the plot becomes a lighter shade of blue), the temperature prediction improves for the mid and high photospheric layers (200-500) km.
 The window that includes the ``knee'' of the spectral line (represented with triangles) yields the best temperature predictions just above the photosphere-chromosphere interface. Here, the ``knee'' of the spectral line is taken to be the apparent wing-core boundary, marked by the vertical orange lines in Fig. \ref{fig:rf} (which for \ion{Ca}{2} K lies at $\lambda - \lambda_0 \sim 0.011$ nm). The ``knee'' of the spectral line seems to mark the separation between two differentiated regimes in terms of the height of maximum sensitivity (HMS) of the spectral line as a function of wavelength.
Above 750 km, the best predictions are attained by the three curves with the lightest shade of green, which include the wavelength ranges in the core of the line proper (from $\sim \lambda_0 - 0.01$ nm all the way to the line center). It seems as if this wavelength range contained the vast majority of the information pertaining to the mid-to-upper chromosphere. The line with the solid circles (which corresponds to the spectral window that includes $\lambda_0$) intersects all other lines and delivers the lowest error of all at the top of the atmosphere.
These results are qualitatively compatible with what we know about spectral line formation: the wings of the line form in the photosphere while the core forms in the high chromosphere \citep[see e.g.][]{Bjorgen2018}.
 
 The results for \ion{Ca}{2} 8542 \AA\ (right) are qualitatively similar, with the core of the line yielding better predictions at higher atmospheric layers and the wing of the line performing better at lower heights. The main difference lies in the prediction error at the very top of the atmosphere (above $\sim 1700$ km), which is significantly worse for IR2 than for the K-line. This is likely due to the fact that the latter forms higher in the atmosphere than the former.

The dashed line represents the errors derived from training the three-layer CNN on the whole spectral line at once. Not surprisingly, it gives the lower bound on the error of the temperature prediction across all spectral windows, and in some height ranges, it performs better than any/all of the individual windows. It is possible that no single narrow wavelength interval contains enough information to capture the LOS integration due to the photon mean-free-path. Also, note that the CNN used to train the whole spectral line uses a larger input and is deeper than the two-layer CNN used to train on short spectral windows, and thus it may have a larger learning ability, leading to slightly smaller errors overall.

The L2-error metric appears to be inversely related to the sensitivity of the spectral line to temperature as a function of height across all of the short wavelength intervals.
In the following section we will compare this CNN-derived sensitivity to the numerical response function of the spectral lines to temperature.




\section{Numerical response functions to temperature}\label{sec:rf}

All spectral lines form over a range of heights in the Sun's atmosphere. Response functions \citep[RF; see, e.g.][]{2016LRSP...13....4D} are defined as the change in the intensity and polarization of a spectral line in response to a small perturbation of an atmospheric parameter in a narrow range of optical depths. RFs provide information about the sensitivity of the Stokes profiles to the atmospheric properties as a function of height, and can help us determine whether a spectral line carries or not (useful) information to diagnose a particular physical parameter at such height.

In this section, we calculate the response functions of \ion{Ca}{2} K and IR2 to temperature following the strategy described in \citet{2016MNRAS.459.3363Q}. Instead of working on an optical depth grid, we work directly on a geometric height grid in order to compare the RFs to the CNN-derived sensitivity obtained for each spectral line in the previous section. 

\begin{figure}[t!]
  \includegraphics[width=0.9\textwidth]{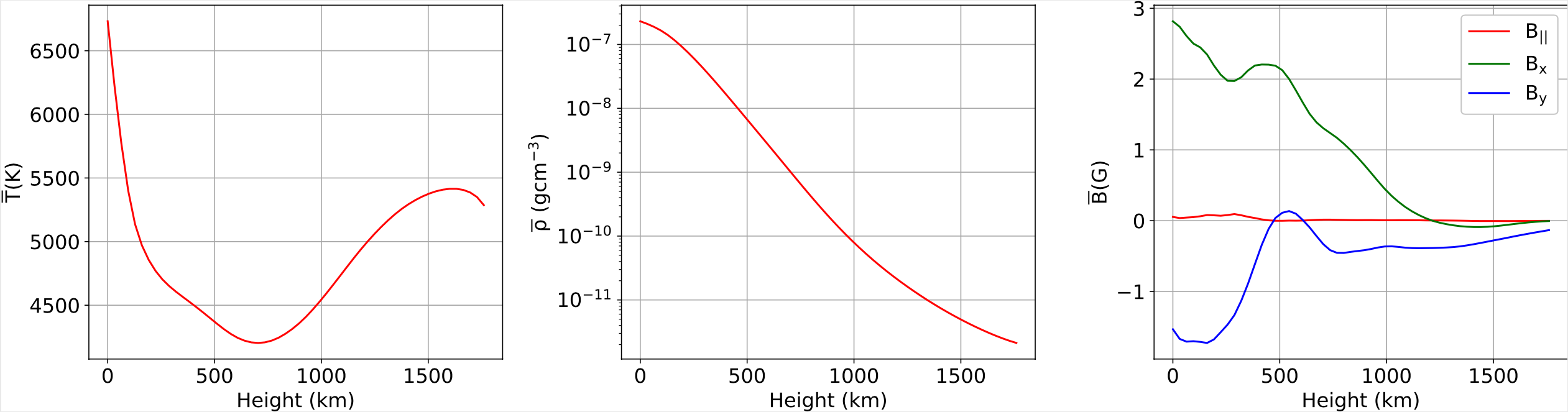}
\caption{Physical properties of the spatially averaged MURaM quiet Sun run.  \label{fig:AvgDTB}}
\end{figure}

RFs are model dependent \citep[][]{2016LRSP...13....4D}. This means that each 1D model atmosphere extracted from the MURaM simulation will result in a slightly different response function to temperature. In order to produce meaningful RFs for the whole dataset, we computed the average  atmosphere from the entire MURaM atmospheric database and used that as the reference model for the calculation. Fig. \ref{fig:AvgDTB} shows the horizontally averaged atmospheric quantities as a function of height. The left panel represents the average temperature ($\overline T$), which drops fast from the bottom of the photosphere towards the temperature minimum (around 700~km), and then rises into the chromosphere. Note that this particular MURaM simulation does not achieve realistic temperatures at the base of the corona, which has an impact on the heights of formation of the spectral lines. 
The middle panel shows the average density ($\overline \rho$), which drops exponentially with height; and lastly, the panel on the right contains the averages of the 3 magnetic field components ($B_\parallel$, $B_x$ and $B_y$, where $\parallel$ refers to the vertical direction, and $x$ and $y$ are in the perpendicular plane). This quiet-Sun MURaM simulation is set up to have a zero vertical net flux ($\overline B_\parallel=0$) and small, yet fluctuating averages of the horizontal components. In this figure, $\overline B_\parallel$ is not strictly zero. This is likely due to the fact that our MURaM dataset comprises just over $\sim 1\%$ of the original snapshot (only 1 in every 8 grid points in each horizontal direction was used to construct the atmospheric database).

For each height $z_i$, we apply a perturbation of $\Delta T=+25$~K to the temperature of the base model and calculate the emergent spectra $I^{+}_i$. We repeat this exercise by applying a negative perturbation of $\Delta T =-25$~K to the base model and computing the corresponding $I^{-}_i$. 
From the emergent intensity spectra, the response of Stokes I to temperature for each wavelength and height is given by:

\begin{equation}
\mathbf{R}_T(\lambda,z_i) = \frac{I^{+}_i -I^{-}_i}{2 I_C \Delta T}
\end{equation}

where $\mathbf{R}_T(\lambda, z_i)$ is the response function, normalized by the local continuum intensity $I_C$. 

\begin{figure}[t!]
\centering
\includegraphics[width=0.45\textwidth]{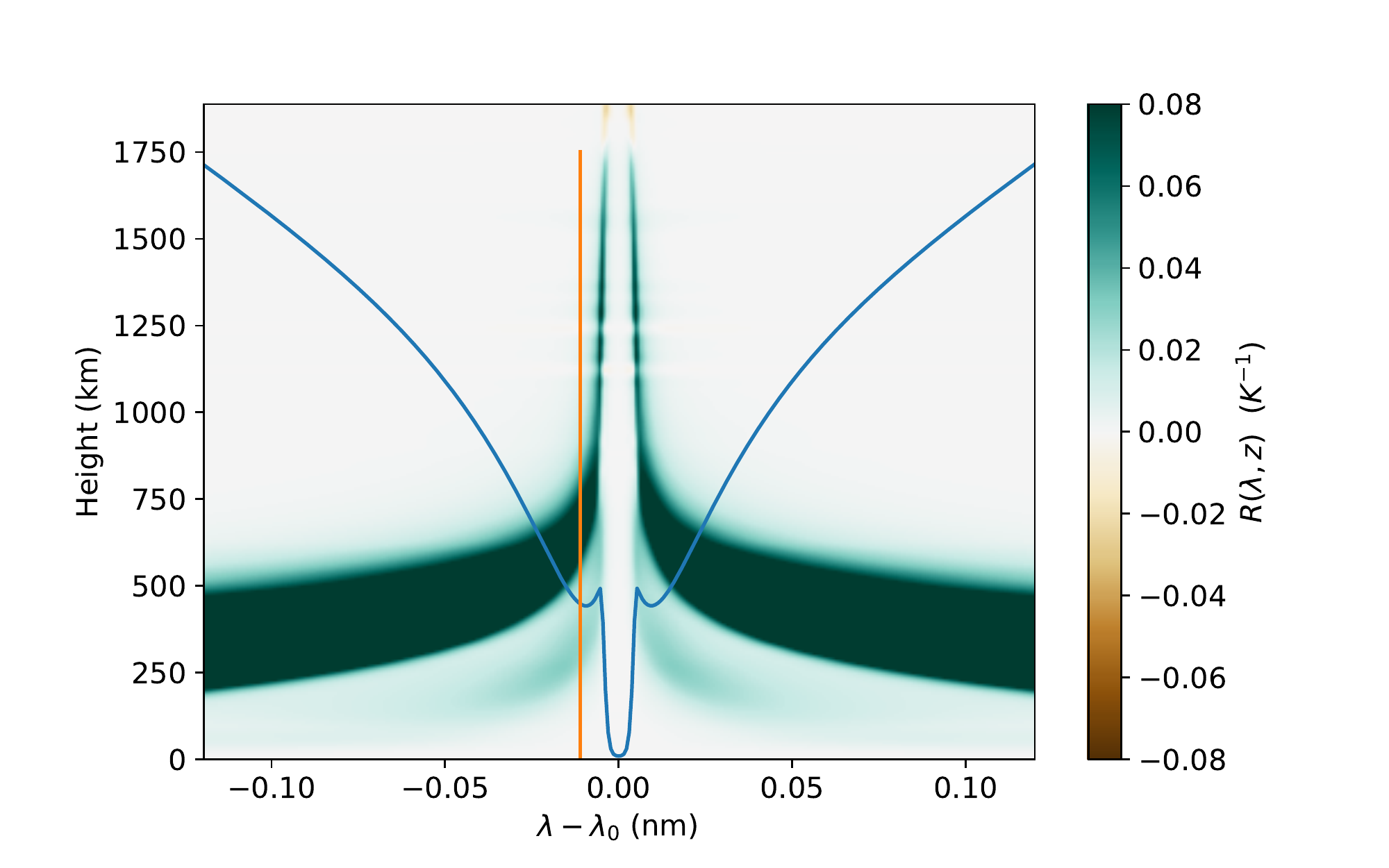}
\includegraphics[width=0.45\textwidth]{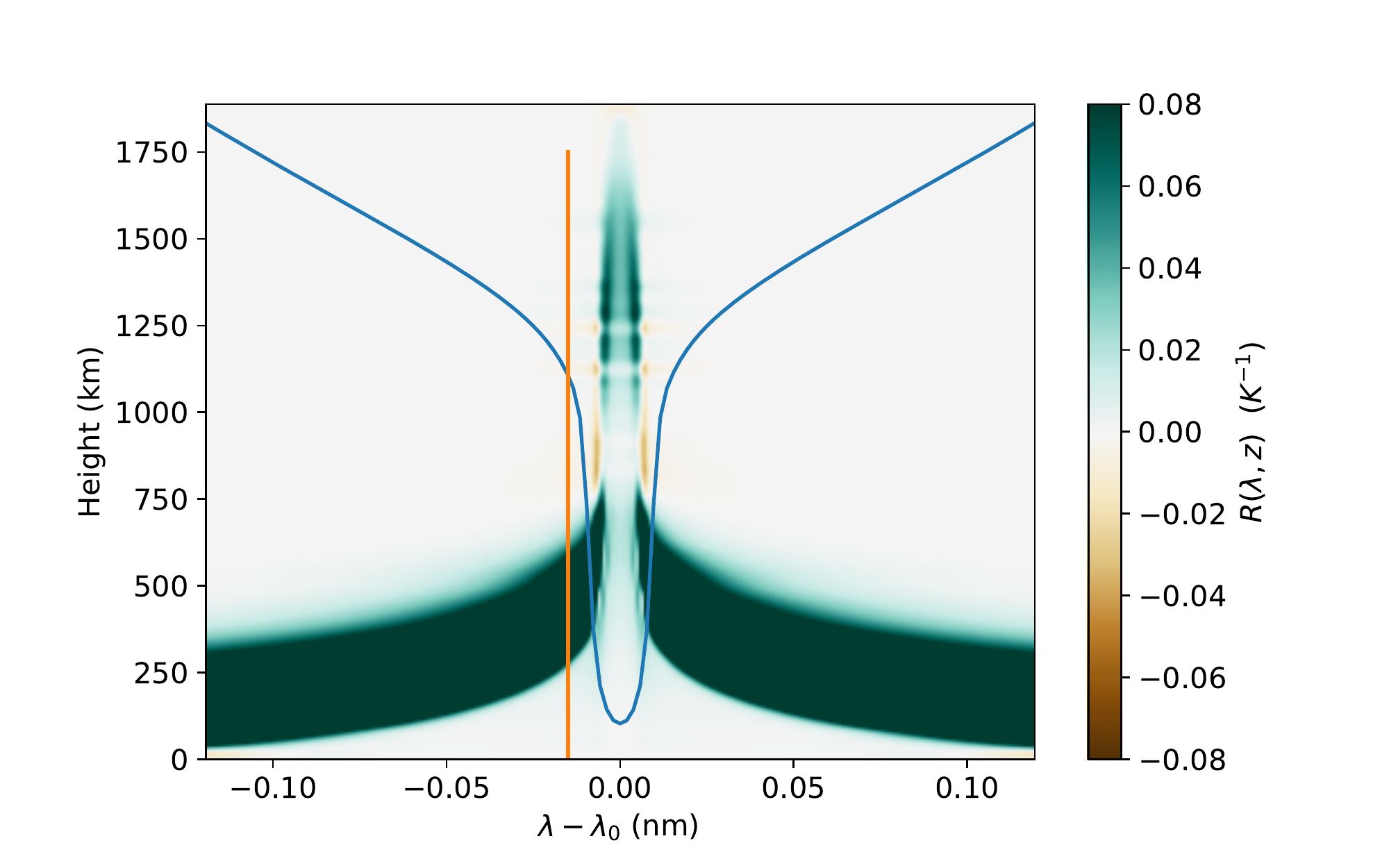}
\caption{Response functions of \ion{Ca}{2} K (left) and \ion{Ca}{2} 8542 \AA\ (right) to temperature in the average MURaM atmosphere. The x-axis represents wavelength and the y-axis geometric height in the mean atmospheric model. A positive response (teal) means that the intensity increases in response to a positive temperature perturbation. The average Stokes I (in blue) is overplotted for reference. The vertical orange line marks the approximate position of the "knee" of the spectral lines. \label{fig:rf}}
\end{figure}

Fig. \ref{fig:rf} shows the response functions of of \ion{Ca}{2} K (left) and IR2 (right) to temperature as a function of wavelength on the x-axis and height on the y-axis. A positive response (teal) means that the intensity increases in response to a positive temperature perturbation. As expected, the wings of both spectral lines are sensitive to changes in temperature at photospheric heights, with a very slow dependence on wavelength. It is not until we reach the knee of the spectral line (marked with an orange vertical line), that a sharp increase in the HMS begins. This change in behavior takes place around $\sim 500-700$ km for both lines. 
It is interesting to note that, in this particular atmospheric model, the sensitivity of the core of the K-line to temperature extends to the very top (and beyond) of the model atmosphere, while the sensitivity of the IR2 line drops off around $\sim 1700$ km. This is qualitatively and quantitatively compatible with the results presented in the previous section, where the L2 error for the CNN trained on IR2 shows a sharp increase above this height. In this context, the L2-error metric of the CNN performance shows an {\em inverse relationship} to the response function of the spectral line to temperature. 

\begin{figure}[t!]
\centering
\includegraphics[width=0.45\textwidth]{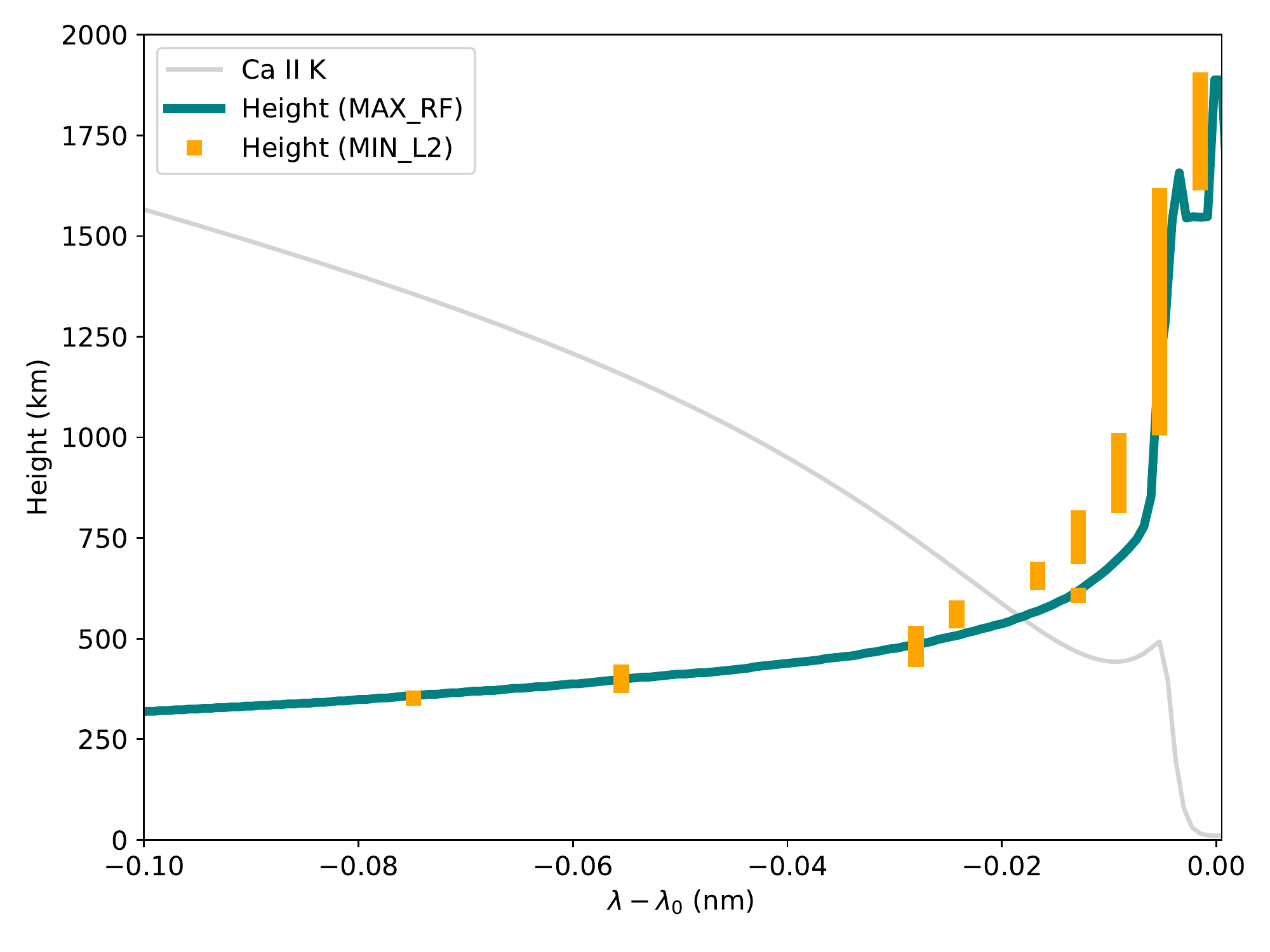}
\includegraphics[width=0.45\textwidth]{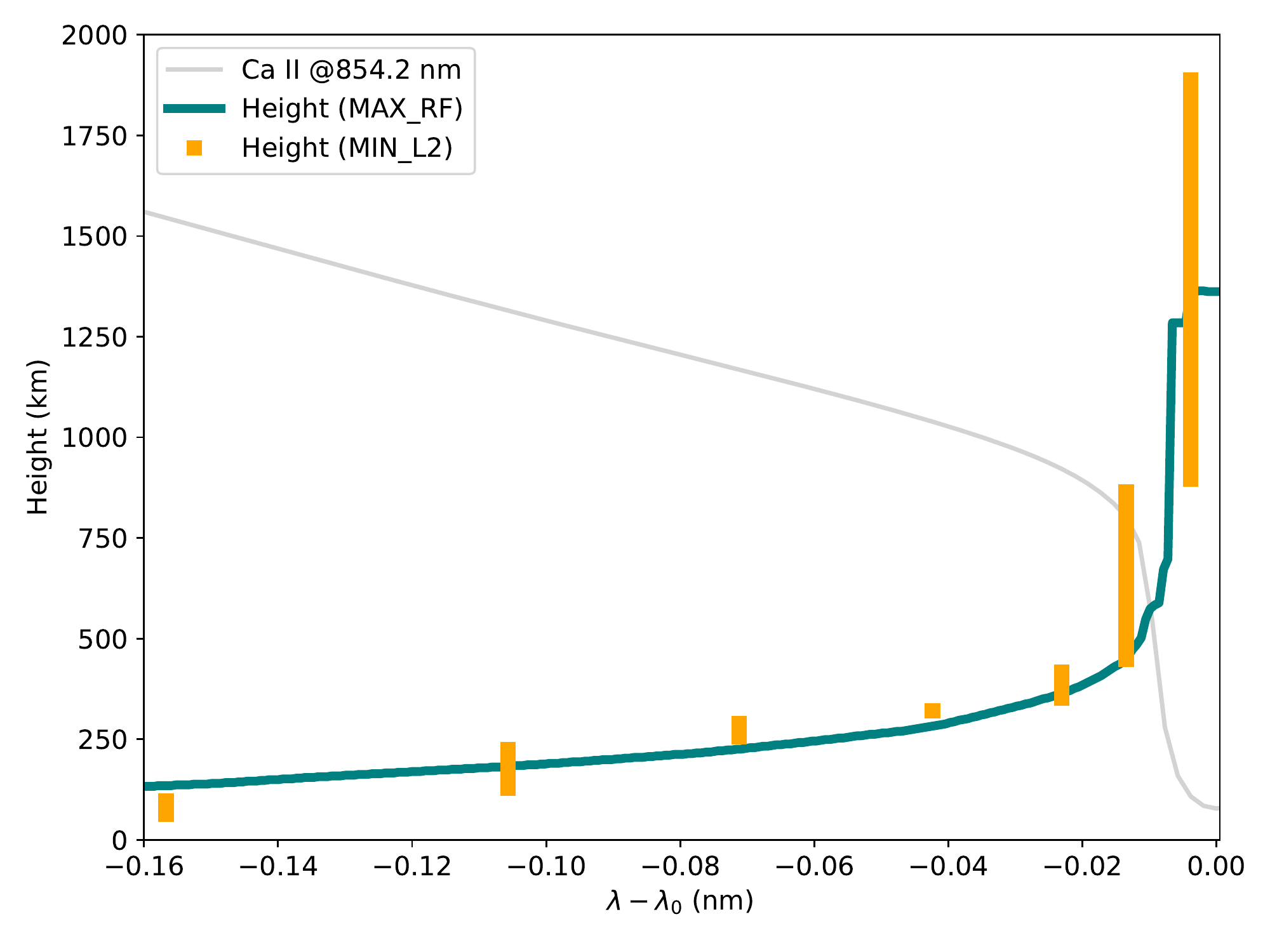}
\caption{Quantitative comparison between the spectral line response function and the L2-error metric of the CNN temperature prediction (\ion{Ca}{2} K on the left and IR2 on the right). In teal we show the wavelength that yields the maximum value of the RF at each height. The orange squares show which wavelength window (represented by its central wavelength) delivers the lowest L2-error for the temperature prediction at each height. The average Stokes I (in grey) is over-plotted for reference. \label{fig:RFvsL2}}
\end{figure}

Figure \ref{fig:RFvsL2} shows a quantitative comparison between these two metrics of spectral line sensitivity. These plots show, in teal, the wavelength that delivers the maximum value of the RF for any given height, or in other words, the height at which the RF peaks as a function of wavelength. We can extract a similar relationship from the L2 curves presented in Fig.  \ref{fig:cnnrf}: at each height in the atmosphere we find the L2 curve (and the central $\lambda$ of the corresponding wavelength interval) that delivers the lowest error in the temperature prediction. This is shown by the orange squares. Note that the orange curve shows a lot more degeneracy than the one derived from the RF, owing to the size of the wavelength intervals on which the CNN was trained. For this reason, a single wavelength interval can yield the lowest L2-error for a range of heights. However, the curves follow each other very well, and seem to tell the same story: 
the HMS shows a monotonic rise as we move from the far wing to the knee of the spectral line; at this wavelength, a dramatic shift in behavior takes place, leading to a sudden increase of this quantity. 

\noindent The agreement between the information contained in the L2-error metric and the RF of the spectral line indicates that the non-linear function learned by the CNN contains physically-meaningful information about the mapping between spectral wavelength and atmospheric height.

\section{Conclusions}

In this work, we analyze the information content learned by a neural network when trained to carry out the inverse mapping between the intensity spectra of two \ion{Ca}{2} lines (separately) and the atmospheric temperature as a function of height.
The numerical setup comprised 36,828 1D model atmospheres extracted from a radiative magneto-hydrodynamic MURaM simulation, and an equal number of \ion{Ca}{2} K and IR2 Stokes I spectra computed with the Hanle-RT code.

First, a three-layer CNN was trained to map the whole spectral line to the atmospheric temperature as a function of height, and its performance was evaluated using an L2-error metric to capture the uncertainty in the temperature prediction (dashed lines in Fig. \ref{fig:cnnrf}). As expected, this prediction error is partially a reflection of the spread of the temperature at each height in the model, but it also captures information about the sensitivity of the spectral line to temperature at different atmospheric heights. This is epitomized by the fact that IR2 leads to much larger prediction errors than the K-line above 1700km, which speaks to the average formation heights of the cores of these two spectral lines in this particular MURaM model.

Then, a two-layer CNN was trained on short wavelength intervals that were slid across the spectral line, from the far wing to the line core. For each spectral window, the CNN is trained to retrieve the temperature as a function of height, and the L2-error metric is collected in each case (the latter is shown by the colored lines in Fig. \ref{fig:cnnrf}). 
The performance of the CNN trained on these short spectral windows shows a monotonic trend in the height that delivers the lowest prediction error. For the lowest atmospheric heights, the window positioned in the far wing of the spectral line delivers the best result (i.e. the smallest error). As we progressively move to higher atmospheric layers, the lowest errors are delivered by spectral windows that are increasingly closer to the core of the line. The best prediction at the top of the atmosphere is achieved only for the window that contains the central wavelength, $\lambda_0$.

Lastly, we compute the numerical response functions to the temperature in order to evaluate the sensitivity of the two \ion{Ca}{2} lines to thermal perturbations in the model. The response function gives us a quantitative measure of the change in Stokes I as a function of wavelength when the temperature is perturbed at any given height in the atmosphere. 
We can draw a one-to-one correspondence between the L2-error metric of the CNN performance (trained on short wavelength intervals) with the response function of the spectral line to temperature. By comparing the wavelength dependence of the height of maximum sensitivity of the RF and the height of minimum L2-error of the CNN temperature prediction, we see a one-to-one mapping between both curves.

This work shows that a NN trained to carry out the inverse mapping between Stokes I profiles and the atmospheric temperature used to generate them, is able to glean information about the wavelength-height sensitivity of the spectral line in a physically meaningful way.


\begin{acknowledgments}
This material is based upon work supported by the National Center for Atmospheric Research, which is a major facility sponsored by the National Science Foundation under Cooperative Agreement No. 1852977. We would like to acknowledge high-performance computing support from Cheyenne (doi:10.5065/D6RX99HX) provided by NCAR's Computational and Information Systems Laboratory, sponsored by the National Science Foundation.

\end{acknowledgments}


\bibliography{rf}{}
\bibliographystyle{aasjournal}

\newpage
\appendix
\section{CNN architecture}\label{appendix:CNN}

Below is the Keras CNN model trained to carry out the mapping between Stokes I and the atmospheric temperature as a function of geometric height. Note that the 2-layer CNN used to train on short spectral intervals has a very similar architecture but is missing the last convolution layer. 

\begin{verbatim}

  import numpy as np
  import tensorflow as tf
  from sklearn.preprocessing import MinMaxScaler
  from sklearn.model_selection import train_test_split
  from tensorflow.keras.layers import Conv2D
  from keras.models import Sequential
  from keras.layers import Dense, Flatten, Activation
  from keras.optimizers import Adam
  from sklearn.model_selection import train_test_split
  
  scaler = MinMaxScaler()
  % X is the input vector, Stokes I and 
  % Y is the output vector which is the temperature at corresponding geometric heights
  scaler.fit(X) 
  X = scaler.transform(X)


  Xx = X.reshape(X.shape[0], X.shape[1], 1, 1)
  Yy = Y.reshape(Y.shape[0], Y.shape[1], 1)
  n_inputs, n_outputs = X.shape[1], Y.shape[1],

  model = Sequential()
  model.add(Conv2D(32, (3,1), strides=(1, 1),activation="elu",padding='same', input_shape=(n_inputs,1,1)))
  model.add(Conv2D(64, (3,1), strides=(1, 1),activation="elu",padding='same'))
  model.add(Conv2D(64, (3,1), strides=(3, 1),activation="elu",padding='same'))
  model.add(Flatten())
  model.add(Dense(n_outputs))
  opt=Adam(lr=0.01, epsilon=None, decay=0.0, amsgrad=False)
  model.compile(loss="mape", optimizer=opt)
  batch_size = 50
  xtr, xte, ytr, yte=train_test_split(Xx, Yy, test_size=0.30, random_state=42)
  model.fit(xtr, ytr, batch_size=batch_size,validation_split=0.03,epochs=150)
\end{verbatim}
\

\section{CNN performance in the presence of LOS velocities}\label{appendix:velocities}
\begin{figure}[ht!]
\centering
\includegraphics[width=0.6\textwidth]{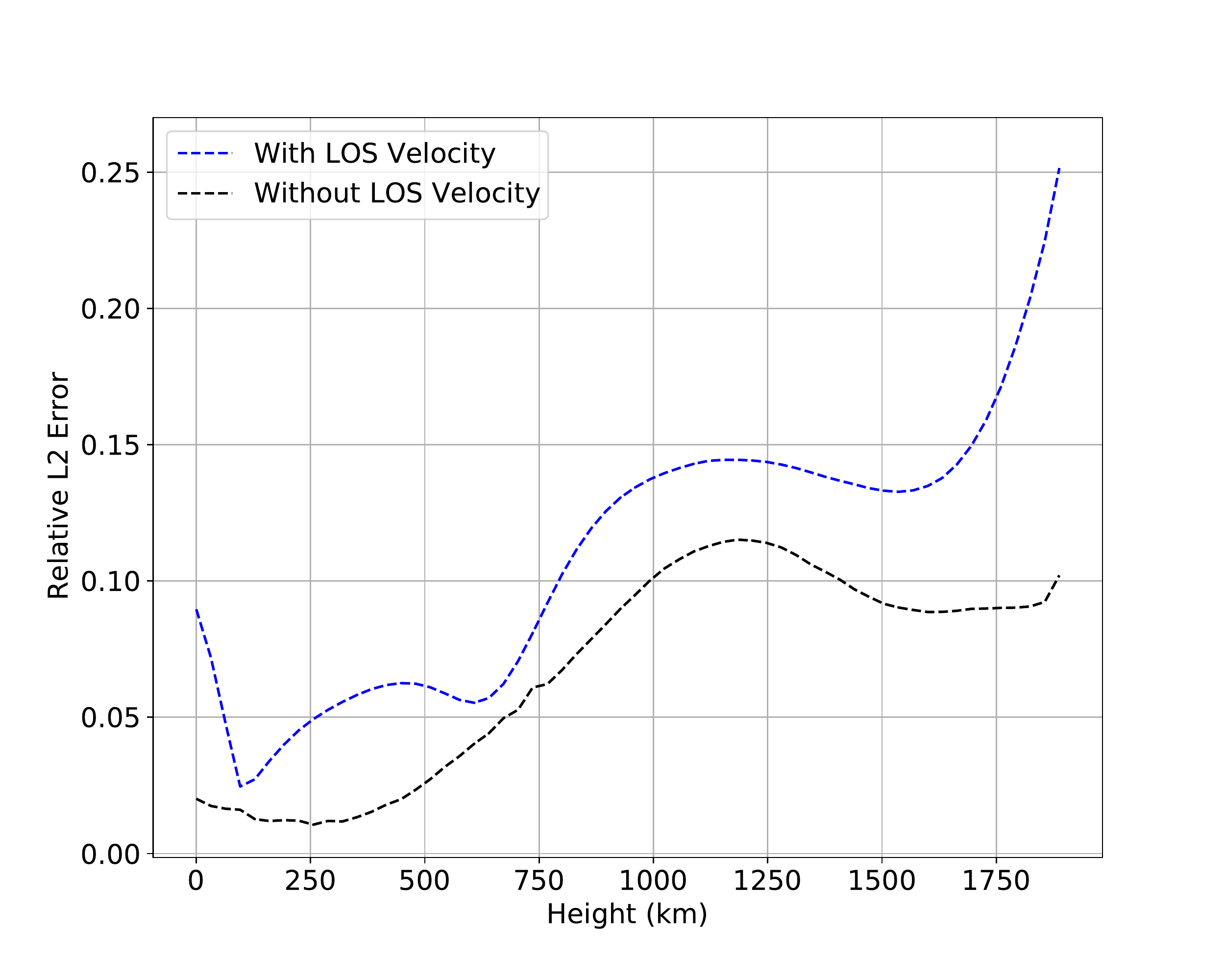}
\caption{Relative L2 error of the temperature retrieval from the CNN inversion of \ion{Ca}{2} K, with and without LOS velocities.\label{fig:L2WithVel}}
\end{figure}
The inversion of spectral lines generated in static atmospheres is of limited practical value. Below we show preliminary results in the accuracy of the temperature retrieval when the Ca {\sc ii} K synthetic spectra are computed in the presence of LOS velocities from the original MURaM simulation. Figure \ref{fig:L2WithVel} shows the relative L2 temperature prediction error in the presence of LOS velocities (blue), which should be compared to the case without velocities (black; this is identical to the black dashed line in the left panel of Fig. \ref{fig:cnnrf}). The two curves present error values within 5\% of each other up to $\sim 1600$km, height above which the L2-error metric for the case with velocities shoots up. This is work in progress and we believe there is room for improvement.




\end{document}